\documentclass[aps,amsmath,amssymb,groupedaddress,twocolumn,nofootinbib,showkeys]{revtex4}
\usepackage{graphicx}

\def\DS {D\!\!\!\!/}

\begin{document}

\title{On the sign problem in dense QCD}  

\author{Stephen D.~H.~Hsu}\email{hsu@uoregon.edu}
\author{David Reeb}\email{dreeb@uoregon.edu}
\affiliation{Institute of Theoretical Science\\University of Oregon, Eugene, OR 97403, USA}

\begin{abstract}
We investigate the Euclidean path integral formulation of QCD at finite baryon density and temperature. We show that the partition function $Z$ can be written as a difference between two sums $Z_+$ and $Z_-$, each of which defines a partition function with positive weights. We call the sign problem \emph{severe} if the ratio $Z_- \, / \, Z_+$ is nonzero in the infinite volume limit. This occurs only if, and generically always if, the associated free energy densities $F_\pm$ are equal in this limit. We present strong evidence here that the sign problem is severe at almost all points in the phase diagram, with the exception of special cases like exactly zero chemical potential (ordinary QCD), which requires a particular order of limits. Part of our reasoning is based on the analyticity of free energy densities within their open phase regions. Finally, we describe a Monte Carlo technique to simulate finite-density QCD in regions where $Z_- \, / \, Z_+$ is small.
\end{abstract}


\date{November 2009}

\maketitle

\bigskip

\begin{center}{\bf 1. Introduction}\end{center}

The Monte Carlo method, which relies on importance sampling, plays a central role in our ability to investigate the nonperturbative properties of strongly coupled quantum field theories. However, many models have a so-called sign problem: we do not know how to express their Euclidean functional integral as a sum over positive quantities, which have a probability interpretation; see, for instance, \cite{signproblemR1,signproblemR2,signproblemR3}. This problem arises in dense QCD, chiral gauge theories and many models of electronic systems. At chemical potentials much larger than $\Lambda_{QCD}$, dense QCD exhibits a Fermi surface, and the effective field theory near the surface has desirable positivity properties \cite{HH1,HH2,HH3,HH4}, however we do not rely on those particular properties here. For an alternative approach to the sign problem using stochastic quantization and complex dynamics, see \cite{Aarts}.

The partition function of a grand canonical ensemble is real, strictly positive and can be expressed as a sum over positive terms (since the Hamiltonian ${\cal H}$, the particle number operator $N$ and all their real linear combinations are Hermitian):
\begin{equation}
\label{grandcanonicalZ}
Z ~=~Z(\beta,\mu)~=~ {\rm Tr} \left( \exp{( - \beta \left( {\cal H} -
      \mu N )\right)} \right)~.  
\end{equation}
Despite these desirable positivity properties, it is difficult
to deal with
$Z$ directly in this form due to the vastness of Hilbert space, over which the
trace is evaluated. It is advantageous to rewrite $Z$ as a Euclidean
functional integral, which sums over \emph{classical} field
configurations rather than quantum states (which are wave functionals
thereof). In the case of dense QCD (Yang-Mills with quarks and
chemical potential): 
\begin{equation}
\label{pathintegral}
Z~=~ \int {\cal D}A ~ \det M(A) ~ e^{- S_G(A)}~,
\end{equation}
where $S_G(A;\,\beta)=\int_0^{\beta} dx^4\int d^3x\, {\cal L}_G^{E} (A)$ is
the thermal Euclidean gauge action. The quark matrix is $M(A)= \DS
\,(A) - m - \mu \gamma_4$, with anti-Hermitian covariant derivative
$\DS \, =\gamma_{\mu} \left(\partial_{\mu} - i g A^a_{\mu} t^a \right)$ and
Hermitian group generators $t^a$ and (Euclidean) $\gamma$-matrices.
For real nonzero chemical potential $\mu$, the eigenvalues of $M$ are complex without grouping into subsets of real and positive products,
and hence $\det M(A)$ is generally complex. The infamous sign
problem arises because importance sampling relies on a positive
definite (in particular, real) measure.

\begin{center}{\bf 2. Two ensembles $Z_\pm$}\end{center}

Nevertheless, from (\ref{grandcanonicalZ}) the partition function $Z$ is
real and can be expressed in a way that makes this manifest. $S_G(A)$ is real and the integral with (Haar) measure ${\cal D}A$ is to be interpreted as a
real Riemannian sum: 
\begin{eqnarray}
Z~&=&~{\rm Re}~Z~=~ \int {\cal D}A ~{\rm Re}\left( \det M(A) ~ e^{- S_G(A)}
\right) \nonumber \\
\label{realZ}
&=&~\int {\cal D}A ~\left({\rm Re} \det M(A)\right) ~ e^{- S_G(A)}~.
\end{eqnarray}
Similar arguments have been made in \cite{Cox:1999nt}.

In the case of dense QCD there is a simple pairing $A \leftrightarrow A'$ of gauge
field configurations which explicitly shows the cancellation of all
imaginary contributions to (\ref{pathintegral}) \cite{Hsu:1998eu}: Since action and measure are gauge-invariant, assume $A_4(x) \equiv 0$ (one could choose any component of $A_\mu (x)$ to be zero; the proof proceeds with only slight modification) for the given gauge configuration $A_{\mu}=A_{\mu}^a t^a$ (or first transform to that
gauge), and define $A_{\mu}' \left(\vec{x},x^4 \right) \equiv
A_{\mu}^* \left(-\vec{x},x^4 \right) $, i.e.~$A_{\mu}'^a
\left(\vec{x},x^4 \right) t^a = A_{\mu}^a \left(-\vec{x},x^4 \right)
\left( t^a \right)^*$. Direct examination of $M(A)$ then shows that
its spectrum $\{ \lambda_i \}$ is mapped to its complex conjugate  $\{
\lambda_i^* \}$ as $A \rightarrow A'$ (the eigenspinors $\psi_i$ of
$M(A)$ are mapped to $\psi_i' \left(\vec{x},x^4 \right) = \gamma_5
\gamma_2 \psi_i^* \left(-\vec{x},x^4 \right)$), so that $\det M(A') =
\left( \det M(A) \right)^*$. Since for the gauge actions
$S_G(A)=S_G(A')$ and for the Haar measures ${\cal D}A={\cal D}A'$, imaginary contributions in (\ref{pathintegral})
cancel pairwise between $A$ and $A'$, leading to (\ref{realZ}) in a
more explicit manner. Note that the above mapping essentially
corresponds to a parity transformation $(\vec{x},x^4) \rightarrow
(-\vec{x},x^4)$ of the fields followed by charge conjugation (complex conjugation with
$\gamma_2$ for fermions). In the lattice formulation of gauge theories, the transformation $U \to U^*$ that maps every gauge link to its complex conjugate already effects a pairing of gauge configurations satisfying $\det M(U^*) = \left( \det M(U) \right)^*$ and $S_G(U^*)=S_G(U)$. A related mechanism works for chiral gauge
theories as well \cite{Hsu:1995qm}. 

\bigskip

Now, in order to deal with \emph{positive} quantities, denote the set of configurations $A$ for which ${\rm Re} \det M(A)$ is positive as $\{ + \}$, and similarly for $\{ - \}$. Then, from (\ref{realZ}),
\begin{eqnarray}
Z &=& \sum_{\{ + \}} \vert {\rm Re} \det M \vert e^{- S_G(A)}
~-~ \sum_{\{ - \}} \vert {\rm Re} \det M \vert e^{- S_G(A)} \nonumber \\
&\equiv& Z_+ - Z_-~.
\label{sum_pm}
\end{eqnarray}
Note, here
\begin{equation}
Z_{\pm} = \sum_{\{ \pm \}} \vert {\rm Re} \det M \vert e^{- S_G(A;\,\beta)}
\label{pm_ensembles}
\end{equation}
are themselves partition functions with positive weights. So, an
analogy with the canonical ensemble in statistical mechanics (at unit temperature) becomes apparent if the Euclidean
action $S_G(A)$ is interpreted as part of the potential energy of a
Hamiltonian $H(A)$ describing a fictitious $4+1$ dimensional theory (as in molecular dynamics methods).
Note, $H(A)$ should not be confused with the Hamiltonian
$\cal{H}$ of the original $3+1$ dimensional system in (\ref{grandcanonicalZ}). 

\begin{figure}[t]
\includegraphics[width=0.95\linewidth]{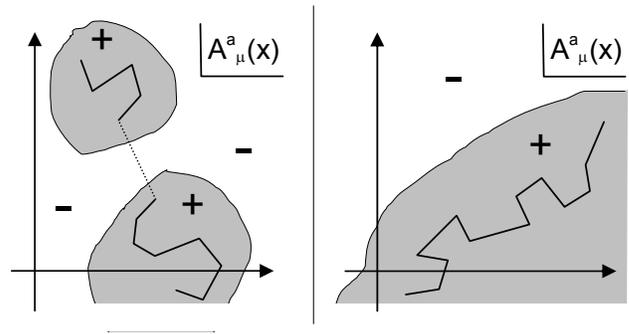}
\caption{Two possibilities for the sets $\{+\}$ (grey) and $\{-\}$ (white). (\emph{left}) One (here the $\{+\}$ set) or both might not be connected; in this case, Hamiltonian evolution will never sample the whole $\{+\}$ set due to infinite potential barriers between the grey and white patches, although molecular dynamics algorithms with finite time step might not see these barriers. (\emph{right}) Each of the two sets $\{\pm\}$ is connected, so sampling of each is possible in principle with molecular dynamics (ergodicity is here assumed).}
\label{hamevolution}
\end{figure}

The fermion determinant in (\ref{pm_ensembles}) can be exponentiated to become a non-local term in this effective
Hamiltonian: $H(A) = V(A)+H_\text{kin}$ with $V(A)=S_G(A) - \ln | {\rm Re} \det M(A) |$. This term is singular ($V(A)=+\infty$) when
${\rm Re} \det M$ passes through zero, which means that a $\{ + \}$
configuration will not evolve into a $\{ - \}$ under Hamiltonian evolution: there is an infinite potential barrier separating the $Z_+$ and $Z_-$ ensembles (\ref{pm_ensembles}) from each other.
Therefore, one can define $Z_\pm$ separately as independent
microcanonical ensembles (using the equivalence of canonical and
microcanonical formulations as in molecular dynamics algorithms; see also next paragraph); and, if $\{ + \}$ as well as $\{ - \}$ is each a connected set (except, possibly, for a subset of measure zero), each ensemble has its own (presumably ergodic) flow (Fig.~\ref{hamevolution}, right panel). That is, modulo this exceptional disconnected measure-zero subset of configurations, typical configurations contributing to $Z_+$ could in principle then be generated by starting with an arbitrary $\{ + \}$ configuration (with appropriate energy; see $E^*_+$ below) and evolving it using the effective Hamiltonian equations of motion for $A$; the same is true for $Z_-$. Whether this procedure for sampling of $Z_\pm$ would work \emph{in practice} on computers depends on the sizes of features in the potential $V(A)$ relative to the time-step size: if the size of the time steps used in the algorithm is not small enough, the infinite potential barriers might not be seen and the algorithm might jump from $\{+\}$ over to $\{-\}$ or over to another disconnected subset of $\{+\}$ (Fig.~\ref{hamevolution}, left).

$Z_\pm$ can be treated similarly to ensembles in statistical mechanics: First
define an ``entropy'' $\Sigma_\pm (E)$ as the logarithm of the number of
$\pm$ configurations $A$ with action (``energy'') $E \equiv S_G - \ln | {\rm Re} \det
M |$. Then both partition functions
\begin{equation}
Z_\pm = \int dE ~e^{-E + \Sigma_\pm (E)} 
\end{equation}
can be expanded about their respective saddlepoints at energies $E_\pm^*$ given by the condition
\begin{equation}
\frac{\partial \Sigma_\pm}{\partial E} \bigg|_{E_\pm^*} = 1~.\label{saddlepoint}
\end{equation}
The saddlepoint energies $E_+^*$ and $E_-^*$ are not necessarily
equal, and, importantly, there is no reason to expect the saddlepoint
values of the exponent to be the same, meaning that, generically, one
of $Z_\pm \approx \exp \left( -E_\pm^*+\Sigma_\pm (E_\pm^*) \right)$
could be exponentially larger than the other. If there is a dominant term 
it must be $Z_+$ since $Z$ is positive. If $Z_+$ is exponentially dominant the situation improves dramatically: it, and therefore practically $Z$, can be evaluated using importance sampling. In this circumstance the sign problem is not severe. Note that dominance by $Z_+$ would not imply the absence of complex phases in the determinant, but only the dominance of contributions with positive real part.

In the following we will make more specific arguments about the relative sizes of $Z_+$ and $Z_-$, also paying closer attention to issues of volume dependence. Although some of our arguments are in principle also capable of supporting exponential discrepancy between $Z_+$ and $Z_-$, we find, in all specific cases considered and through one more general argument, strong evidence that in all open regions of the QCD phase diagram they become equal at infinite volume, or are at least not exponentially separated, thereby constituting a severe sign problem in (at least) all of those open regions. This is in contrast to the expectations based on the saddlepoint approximation in (\ref{saddlepoint}).

\begin{center}{\bf 3. Free energies and sign problem}\end{center}

Without invoking the saddlepoint method, we can define free energy densities $F_\pm$ via\footnotemark[1]\footnotetext[1]{$F_{\pm}(\mu,\beta)$ are the densities corresponding to the \emph{extensive} part of the free energies, formally defined as follows: partition functions $Z$ and $Z_{\pm}$ are finite only for finite volumes $V$, and the volume dependence of the associated densities $f$, $f_{\pm}$, defined via $Z(\mu,\beta,V)=\exp{\left(-V f(\mu,\beta,V)\right)}$, vanishes only in the thermodynamic limit: $F(\mu,\beta) \equiv \lim_{V\to \infty} f(\mu,\beta,V)$. Nevertheless, for finite $V$, at which simulations are done, the volume dependence $f(V)=F+\hat{f}(V)$ can be important and subtle: since zero-density QCD has positive measure, $F_-(\mu=0,\beta)=+\infty$ (or is ill-defined), but for any chosen (not too small) constant $c<\infty$ there seem to be infinite sequences $\mu_n \to 0$, $V_n \to \infty$ such that $f_-(\mu_n,\beta,V_n) \equiv c$ for all $n$; so the order of limits is important. Also, if the volume-dependent pieces $\hat{f}_\pm(V)$ vanish more slowly than $1/V$ in the thermodynamic limit, one can, contrary to naive expectation, have $Z_-/Z_+ \to 0$ in the case $F_+=F_-$, thereby unexpectedly alleviating the severity of the sign problem (but in the case $F_+<F_-$ one always has $Z_-/Z_+ \to 0$, i.e.~no severe sign problem).}
\begin{equation}
Z_\pm \equiv \exp \left( - V F_\pm ( \mu,\,\beta ) \right)~,
\end{equation}
where $V$ is the Euclidean 4-volume in which the theory lives (or the physical lattice volume). In the previous version of this paper \cite{previousversionv2}, we conjectured that
in the large volume limit the generic situation is
$$
F_+ < F_- ~,
$$
which would imply that $Z_+$ exponentially dominates $Z_-$ at large volume $V$. The alternative is that they are exactly equal: $F_+ = F_-$ (recall $Z > 0$, so $F_+ > F_-$ is not possible). It is in this case that the sign problem can be \emph{severe} (in the sense defined in the abstract): the ratio of $Z_-$ to $Z_+$ then generically approaches some nonzero constant (possibly 1) at large volume\footnotemark[1]. In the thermodynamic limit, free energy densities are analytic except at phase boundaries; thus, if $F_+ = F_-$ in some open region, they must be equal {\it everywhere} within the whole intersection of the respective phase regions of $Z_+$ and $Z_-$ (i.e., except, perhaps, at or across phase boundaries where free energy densities might not be analytic). At $\mu = 0$ we know $Z_- = 0$, i.e.~$F_- = +\infty$, and $Z_+=Z>0$, so it seemed natural to us to assume that there would be an open region of small $\mu$, extending into the $(\mu,\beta)$ plane, where $F_+ < F_-$.

It turns out that this last assumption is most likely incorrect. Results which we will discuss below strongly suggest that even at arbitrarily small (but nonzero) $\mu$, $Z_+$ does \emph{not} exponentially dominate $Z_-$. That is, $F_+ = F_-$ exactly, even at small nonzero $\mu$. The order of limits is important\footnotemark[1]: taking $V$ to infinity for fixed nonzero $\mu$ (no matter how small) leads to large sign fluctuations. Ordinary QCD is only obtained by taking $\mu \rightarrow 0$ before taking $V \rightarrow \infty$. Strangely, the case $\mu=0$ we are most familiar with is the atypical one!

Although the logic based on (piecewise) analyticity of $F_\pm$ inside their respective phase regions is correct, what we originally \cite{previousversionv2} conjectured to be ``exceptional'' regions in the phase diagram are the typical ones, and vice versa. There seems to be no open region at small $\mu$ where $Z_+$ dominates, rather only the axis where $\mu$ is exactly zero\footnotemark[2]\footnotetext[2]{In \cite{SV1}, the authors find a particular order of limits which results in no sign problem for $T$ exactly equal to zero and small $\mu$ ($\mu < m_{\pi}/2$). However, this again is not an open region in the phase diagram to which analyticity arguments would apply, and it also depends on the order of limits.}.

\begin{figure}[t]
\includegraphics[width=0.95\linewidth]{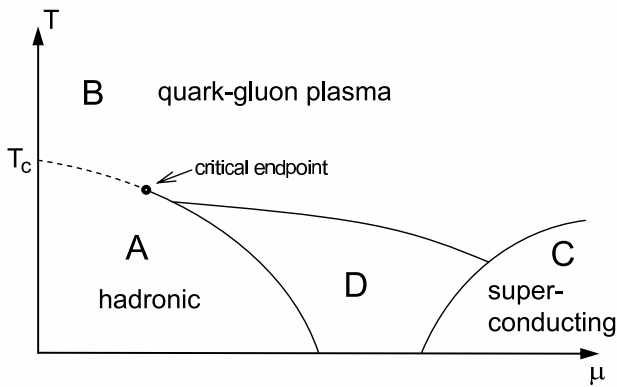}
\caption{Schematic phase diagram of QCD. We argue that dense QCD has a severe sign problem almost everywhere in the $(\mu,~T)$-plane. Since within a particular intersection of a phase of $Z_+$ and a phase of $Z_-$ the free energy densities $F_\pm$ are both analytic, we only have to argue that $F_+=F_-$ in some open patch of that intersection to show equality in the entire phase intersection. The phases of the theory $Z$, shown here in the diagram, might not or might only partially coincide with the phases of $Z_+$ and $Z_-$ (see esp.~discussion of hadronic phase in Section 4). For realistic quark masses, there probably is a smooth crossover between phases A and B, implying analyticity of the free energy density of $Z$ across this boundary. Whether a separate phase D of exotic nuclear matter exists is speculative.}
\label{freeenergy}
\end{figure}

In fact, one can probably go further and argue that (in the large volume limit) $F_+ = F_-$ almost everywhere in the QCD phase diagram, meaning that the sign problem is, in a sense, maximally severe. The analytic properties of $F_\pm$ are useful in deducing this result, since one only has to find an open subregion in each phase of $Z_+$ and $Z_-$  where the sign problem is severe -- see Fig.~\ref{freeenergy}. It might not be enough, though, to find one such open region in each phase of $Z$ (see the examples of hadronic, quark-gluon plasma and color superconductor phases in Section 4) to argue in this way for (or against) a severe sign problem in the whole respective phase of $Z$ since $F_+$ and $F_-$ might have more (or at least different) phases and phase boundaries than $Z$ (see esp.~discussion of the hadronic phase in the following section). Note, in the figure we have allowed for the possibility of an exotic dense nuclear phase in region D; whether such a phase exists is speculative.

Besides applying the above reasoning based on analyticity properties of the free energy densities $F_\pm$ to specific regions of the QCD phase diagram and finding a \emph{severe} sign problem -- in the sense defined above -- in each region (as we will do in Section 4), there exists one other general argument that the sign problem might be \emph{strong} (in a possibly somewhat different sense, namely as measured by the volume dependence of sign or phase averages in some appropriate or chosen ensembles) almost everywhere on the phase diagram: each such sign or phase average can be written as the ratio of two (positive) partition functions, which at finite volume $V$ have associated free energy densities, e.g.
\begin{eqnarray}
\langle\text{sgn}\rangle_{\vert\text{Re}\det\vert}&=&\frac{\int{\cal D}A\,\text{Re}\det M\,e^{-S_G}}{\int{\cal D}A\,\vert\text{Re}\det M\vert\,e^{S_G}}=\frac{Z}{Z_{\vert\vert}}\nonumber\\&=&\frac{e^{-VF}}{e^{-VF_{\vert\vert}}}~.\label{signaverage}
\end{eqnarray}
Since the average sign is no greater than 1, we have $F\geq F_{\vert\vert}$ necessarily, and the case $F=F_{\vert\vert}$ seems to be the exception, especially if the physics described by the two partition functions (or, equivalently, by the two free energy densities) is different. In this line of reasoning, it is often argued that the sign problem is strong, since sign averages vanish exponentially fast in the space (or lattice) volume $V$, as soon as there exist configurations $A$ on which the two partition functions disagree; i.e., in the example (\ref{signaverage}), if there exist configurations with $\text{Re}\det M(A)\neq\vert\text{Re}\det M(A)\vert$, meaning $Z_-\neq0$. Similar kinds of reasoning can be applied to other sign and phase averages, like the commonly considered $\langle e^{i\,\text{arg}(M)}\rangle_{\vert\det\vert}$ or $\langle e^{2i\,\text{arg}(M)/N_F}\rangle_{\vert\det\vert}$ \cite{SV,SV1} (with $N_F$ the number of quark flavors), etc. Note that, in our notation (\ref{sum_pm}), the specific average (\ref{signaverage}) equals $\left(Z_+-Z_-\right)/\left(Z_++Z_-\right)$, suggesting that $Z_-/Z_+$ approaches 1 (for large volumes $V$) at most points in the phase diagram, thereby supporting the claim from the previous paragraph about the severity of the sign problem in dense QCD.

\begin{center}{\bf 4. Phase diagram}\end{center}

The most powerful techniques we are aware of are those of Splittorff and Verbaarschot \cite{SV,SV1}. 
Let $\theta=\theta(A)$ be the phase of the determinant of the quark matrix $\hat{M}=\hat{M}(A)$ of a single flavor:
\begin{equation}
e^{2 i  \theta} = \frac{\det \hat{M}}{\det {\hat{M}}^*}~.
\end{equation}
(Then, for a theory with $N_F$ flavors, the full fermion determinant $\det M = \det \hat{M}^{N_F}$ has phase angle $N_F \theta$.) In \cite{SV1} it is shown, using chiral perturbation theory, that in the low temperature and density region (lower left corner of region A in the diagram of Fig.~\ref{freeenergy}) the distribution $\rho ( \theta )$ governing the angle $\theta$ (viewed as a non-compact variable $\theta \in \left(-\infty,+\infty\right)$) is Gaussian and has a width which grows as $\sqrt{V}$.\footnotemark[3]\footnotetext[3]{Note, these results are obtained in \cite{SV1} by considering phase distributions $\rho(\theta)$ computed with respect to the phase-quenched measure $\vert\det M(\mu)\vert$, where $M(\mu)$ is the fermion matrix with quark chemical potential $\mu$; this phase-quenched measure is identical to the measure of quarks coupled to an isospin chemical potential $\mu_I=\mu$. Nevertheless, since here and in \cite{SV1} $\rho(\theta)$ is the distribution of the phase of the determinant of the fermion matrix with \emph{quark} chemical potential, our results (and the results in \cite{SV1}) do apply to dense QCD with quark chemical potential, as claimed.} This means that in the thermodynamic limit $V\to \infty$ both signs of ${\rm Re} \det M$ (i.e., both $Z_+$ and $Z_-$) contribute significantly to $Z$, even in such a way that $Z_-/Z_+ \to 1$ as $V\to \infty$ in this patch of region A. To obtain ordinary (zero density) QCD, one must take $\mu$ to zero before taking $V$ to infinity\footnotemark[1]. Our analyticity arguments from above then suggest that there is a severe sign problem in the (open) intersection of the two phases of $Z_+$ and $Z_-$ which both contain the patch of low $\mu$ and $T$. Note that this intersection is most likely smaller than the whole phase region A: both $Z_+$ and $Z_-$ seem to have a phase transition (``onset of pion condensation'') at $\mu$ around $m_\pi/2$ (half the pion mass) in the temperature range of phase A; the phase transition cancels out in the difference $Z=Z_+-Z_-$. For $T=0$ this phase transition at $\mu=m_\pi/2$ is known as the Silver Blaze transition \cite{silverblazeR1,silverblazeR2}, and at $T>0$ in the hadronic phase there is growing numerical and theoretical \cite{splittorffprivate} evidence for a very similar transition in $\mu$. (Note, the sign problem seems to be even more severe to the right of this transition line.) If the finite temperature, low density phase transition (dashed line in Fig.~\ref{freeenergy}) is a crossover in the theory $Z$, as is currently believed to be the case for realistic quark masses, there might also only be a smooth crossover for the $Z_\pm$ ensembles across this line, and this would suggest that right across the crossover there is also a severe sign problem in an open subset of the quark--gluon plasma phase B.

Next, examining this phase B separately, consider the following two partition functions.
Let $Z_{N_F}$ describe dense QCD with an even number of flavors $N_F$, and let $Z_{| N_F |}$ describe a model with $N_F/2$ flavors with chemical potential $\mu$ and $N_F/2$ flavors with opposite potential $- \mu$. (All quarks are assumed to be degenerate.) In $Z_{| N_F |}$ the phase factors cancel and only the magnitude of the determinant appears in the functional integral. Clearly, from the path integral representation,
$$
Z_{N_F}\leq Z_{| N_F |}~.
$$
Now write
\begin{equation}
\label{ratio}
\frac{Z_{N_F}}{Z_{| N_F |}}  = \frac{\exp{( - VF_{N_F})} } { \exp{( - VF_{|N_F|}})}~.
\end{equation}
Unless the two free energy densities in (\ref{ratio}) are exactly equal, which would be peculiar since the two models have very different physics, this ratio will be zero in the thermodynamic limit $V\to\infty$. Equality of $F_{N_F}$ and $F_{| N_F |}$ would require that the free energy of QCD be a function only of even powers of individual chemical potentials for each flavor. In the quark--gluon plasma phase (large temperature and small $\beta \mu$; upper left corner of region B in Fig.~\ref{freeenergy}) one can compute the free energy density in perturbation theory \cite{Vuorinen:2003fs,SV1}, and indeed terms linear in individual chemical potentials (e.g., $\sim \left(\sum \mu_f\right)^2\,$) are found, so the ratio (\ref{ratio}) is zero at infinite volume. We expect this to be the case generically, on physical grounds. 

Now, eqn.~(\ref{ratio}) can be rewritten using the distribution function $\rho (\theta)$ (computed in the phase quenched theory) as
\begin{equation}
\label{rho}
\frac{Z_{N_F}}{Z_{| N_F |}}  = \int d\theta~ \rho ( \theta ) \cos ( N_F \theta )~,
\end{equation}
from which it is clear that the vanishing of this ratio implies either severe sign oscillations in $Z_{N_F}$ (i.e.~the distribution $\rho(\theta)$ is flat such that both signs of $\cos (N_F\theta) \sim {\rm Re}\det M$ contribute significantly to $Z_{N_F}=Z_+-Z_-$, even implying $Z_- \, / \, Z_+ \to 1$), or that $\rho (\theta)$ becomes arbitrarily peaked at values $\theta$ where $\cos (N_F \theta)$ vanishes. This latter possibility requires that $\det M$ becomes purely imaginary for essentially all gauge backgrounds, which seems implausible.

Thus, our analyticity reasoning, combined with other inputs, provides strong evidence for a severe sign problem in the part $\mu<m_\pi/2$ of the hadronic phase A (small $\mu,T$; extended to all of this subregion of A by analyticity) and in the small $\mu$ part of the high temperature phase B (extendible by analyticity to other regions as long as $Z_\pm$ are both analytic). Note, in phase B the order of limits seems to be important as well: if, at \emph{fixed} volume $V$, the temperature is sent to infinity $T\to+\infty$, the sign problem seems to become arbitrarily \emph{weak}. This is supported by numerical lattice simulations and by the theoretical observation that in the infinite temperature limit (in lattice regularization) only the trivial plaquette configuration contributes with significant Boltzmann weight \cite{mplombardoprivate}.

Finally, we consider the low temperature, high density phase (region C in Fig.~\ref{freeenergy}), which is known to exhibit color superconductivity \cite{cscR1,cscR2,cscR3,cscR4,cscR5}. This phase has a low energy effective field theory description (of modes near the Fermi surface) which has good positivity properties \cite{HH1,HH2,HH3,HH4}. However, that does {\it not} imply that the sign problem is absent in the usual Euclidean path integral (\ref{pathintegral}). In fact, if there were no severe sign problem -- i.e., $Z_+$ would dominate $Z_-$ exponentially at large volume -- QCD inequalities \cite{ineqR1,ineqR2,ineqR3,ineqR4,ineqR5,ineqR6} would apply to this phase. Such inequalities forbid the spontaneous breaking of vector symmetries such as baryon number, which is in contradiction to explicit calculations in the dense phase \cite{HH2}. This suggests that the part of the high density (CFL) phase where those calculations are applicable (large $\mu$, small $T/\mu$), again extended to the common analyticity region of $Z_+$ and $Z_-$, also has a severe sign problem.

Since the very existence of a separate exotic nuclear phase (region D in Fig.~\ref{freeenergy}) is speculative, what we can infer about it is very limited. Two comments are perhaps in order: (1) if the phase D is not separated by a phase boundary in the $Z_\pm$ ensembles from the regions A, B or C (i.e., if there is only a smooth crossover), then analyticity would suggest a severe sign problem there as well. (2) unless the free energy in region D is independent of the signs of the individual chemical potentials for each flavor (i.e., a function only of even powers of the $\mu_f$'s), then the argument we used for region B would also apply, indicating a severe sign problem.

\begin{center}{\bf 5. Degrees of severity of the sign problem}\end{center}

In this paper we have defined the sign problem as \emph{severe} whenever the ratio $Z_- \, / \, Z_+$ is nonzero in the thermodynamic limit. The general argument presented at the end of Section 3 and the specific cases considered in the previous section suggest that this may be the case in almost the entire phase diagram. Note, however, that the precise value of $Z_- \, / \, Z_+$ at infinite volume also depends on the subleading (in volume) behavior of the free energy densities\footnotemark[1]. Some may prefer to call the sign problem severe only when $Z_-/Z_+\to1$ as $V\to \infty$. It is possible that regions exist where, by this latter criterion, the problem is not severe: that is, where $Z_+ \gg Z_-$ at large volume. One candidate region might be at low density and zero temperature ($\mu < m_\pi / 2$ and $T\to 0$ as $V\to \infty$ in a certain order of limits), based on analytical results in \cite{SV,SV1}.

Another region with relatively mild sign problem might be around the high temperature phase transition line extending from $(T_c,\,\mu=0)$ to $\mu>0$ (dashed line in Fig.~\ref{freeenergy}), based on lattice results in \cite{fodorkatzR1,fodorkatzR2,fodorkatzR3,handsphasetransitionR1,handsphasetransitionR2,imaginarymuR1,imaginarymuR2,imaginarymuR3,imaginarymuR4}, or around the critical point itself. The fact that the Taylor expansion with coefficients computed at $\mu = 0$ \cite{handsphasetransitionR1,handsphasetransitionR2} or at imaginary $\mu$ \cite {imaginarymuR1,imaginarymuR2,imaginarymuR3,imaginarymuR4} agrees with results from multi-parameter reweighting at nonzero $\mu$ \cite{fodorkatzR1,fodorkatzR2,fodorkatzR3} in this region might suggest that the sign problem could be relatively mild there, at least for the lattice volumes considered. The three methods start to disagree at large $\mu \gtrsim 1.3 \, T_c$ and do so significantly in the region near the QCD critical point \cite{deForcrand:2006ecR1,deForcrand:2006ecR2,deForcrand:2006ecR3}.

Let us consider the (multi-parameter) reweighting method explicitly. In this method one computes
\begin{equation}
\label{rw}
Z( \mu, \beta ) = \left\langle
\frac{e^{- S_G(\beta)} \det M( \mu)}{ e^{-S_G( \beta_0 )} \det M (\mu_0 )}
\right\rangle_{\left(\mu_0, \beta_0\right)}
\end{equation}
using configurations generated from the $( \mu_0, \beta_0)$ ensemble (${\rm Re}\,\mu_0 = 0$ to enable importance sampling). It is assumed that these configurations have significant ``overlap'' with ``typical'' configurations of the $( \mu, \beta )$ target model. The reweighting factor in angle brackets links the two ensembles. In case of a severe sign problem, the reweighting method faces the following challenges \cite{challenges}: (1) there are no ``typical'' configurations in the target theory and there is no meaning to a good overlap between typical $(\mu_0, \beta_0)$ and target configurations, (2) $Z ( \mu, \beta )$ must be exponentially smaller than $Z ( \mu_0, \beta_0 )$ and is therefore sensitive to error; equivalently, the reweighting factor, whose average itself is essential if one wants to compute an operator average $\langle {\cal O} \rangle$, must exhibit large sign fluctuations and numerically subtle cancellations \cite{Ejiri:2004yw,rewfactor}. Thus, a relatively weak sign problem seems to be a necessary condition for reweighting methods to work.

In the remainder of the paper we describe a method of simulating dense QCD (see also \cite{rewfactor}) which will work in this case of a weak sign problem as well, with probably some advantages.

\begin{center}{\bf 6. A Monte Carlo method at $\mu \neq 0$ for $Z_+ \gg Z_-$}\end{center}

At a point in the phase diagram where $Z_+$ dominates $Z_-$ (either exponentially or just by a large enough factor\footnotemark[4]\footnotetext[4]{The method might also work at \emph{finite} simulation volume $V$ even if $Z_- / Z_+ \to 1$ in the thermodynamic limit, because subleading contributions to the free energy densities could render $Z_+ \gg Z_-$ for sufficiently small volumes $V$,\footnotemark[1] although in this case finite volume effects might be so important that one might not actually be simulating the desired (infinite volume) system. On the other hand, rather than assessing the validity of a simulation by looking at finite-size effects in the ratio $Z_-/Z_+$, one should probably compare the correlation lengths of the physics inside the simulated volume to the lattice size itself, regarding a simulation as reliable if the latter is (much) larger than the former.}), we can evaluate any dense QCD correlator in good approximation using the new partition function
\begin{equation}
Z_{\rm new} \equiv Z_+ + Z_-
\end{equation}
instead of the original $Z$ (cf.~(\ref{sum_pm})). By assumption, $Z_{\rm new}$ differs from $$Z = Z_+ - Z_-$$ by only a small amount ($Z / Z_{\rm new} \simeq 1$) since $Z_+ \gg Z_-$. Therefore, a lattice simulation employing the partition function $Z_{\rm new}$ yields almost the same results as a simulation using $Z$, which is the desired partition function to simulate were it not for its sign problem. But now the functional integral
\begin{equation}
Z_{\rm new} = \int {\cal D}A ~\vert {\rm Re} \det M(A) \vert~ e^{-S_G(A)}
\label{Znew}
\end{equation}
has positive weights and allows for importance sampling. 

To show more concretely how this change $Z \rightarrow Z_{\rm new}$ enables importance sampling, we can reformulate in terms of expectation values of a real observable{ ${\cal O}(A)$:
\begin{eqnarray}
\langle {\cal O} \rangle ~&=&~ {\rm Tr} \left( {\cal O} \, e^{ - \beta \left( {\cal H} - \mu N \right)} \right) \nonumber \\
&=&~ \frac{1}{Z} \int {\cal D}A ~ \det M(A) ~ e^{- S_G(A)} ~ {\cal O}(A)~.
\end{eqnarray}
From the representation as a trace, the expectation value is clearly real if ${\cal O}$ is Hermitian. Thus, by applying the same reasoning leading from (\ref{pathintegral}) to (\ref{realZ}), one obtains
\begin{equation}
\langle {\cal O} \rangle ~=~ \frac{1}{Z} \int {\cal D}A ~ \left( {\rm Re} \det M(A) \right) ~ e^{- S_G(A)} ~ {\cal O}(A)~.
\label{realZO}
\end{equation}
For operators that are invariant under the above parity-charge conjugation operation, i.e.~${\cal O}(A)={\cal O}(A')$, this can also be justfied by the pairing arguments $A \leftrightarrow A'$ from Section 2. Then, following the equivalent of (\ref{sum_pm}) to (\ref{Znew}), one arrives at
\begin{equation}
\langle {\cal O} \rangle ~\approx ~\frac{\int {\cal D}A ~ \vert {\rm Re} \det M(A) \vert ~ e^{- S_G(A)} ~ {\cal O}(A)}
                 {\int {\cal D}A ~ \vert {\rm Re} \det M(A) \vert ~ e^{- S_G(A)}}~,
\label{MCsampling1}
\end{equation}
in which the approximation is good assuming that $Z_+ \gg Z_-$ and that ${\cal O}$ does not give exponentially large contributions on the $\{ - \}$ set of configurations $A$. (This last condition would be satisfied by, e.g., any polynomial in the gauge field.) (\ref{MCsampling1}) becomes an exact equation (without any overhead in numerical implementations) if the sign of $\text{Re}\det M$ in the numerator and denominator is taken into account (on the actually sampled configurations). Then, the method is very similar to other reweighting methods:
\begin{equation}
\langle {\cal O} \rangle ~= ~\frac{\int {\cal D}A ~ {\rm Re} \det M(A)  ~ e^{- S_G(A)} ~ {\cal O}(A)}
                 {\int {\cal D}A ~ {\rm Re} \det M(A) ~ e^{- S_G(A)}}~.
\label{MCsampling}
\end{equation}
Conventional importance sampling with Monte Carlo techniques can be used to compute the expectation value of observables -- the weights $\vert \text{Re}\det M\vert e^{S_G(A)}$, with which the sampling in (\ref{MCsampling}) is done, are positive and can be interpreted as probabilities\footnotemark[5]\footnotetext[5]{Note, if one generates typical configurations using microcanonical methods -- i.e., relying on ergodic Hamiltonian flow -- it is necessary to modify the fermionic term in the effective Hamiltonian to allow for $\{ + \}$ modes to evolve to $\{ - \}$ modes and vice versa. One possible modification is $\ln ( \vert {\rm Re} \det M \vert + \epsilon )$, where $\epsilon$ is some small positive quantity. This smoothes out the singularity in the action at zero real part of $\det M$ and allows the ergodic flow to generate configurations of either sign. Alternatively, one might rely on some intentional randomness (possibly due to finite step-size on a computer) in the generation of new configurations or on sufficiently big numerical inaccuracies; see, e.g., \cite{whitewilkins}. See also Fig.~\ref{hamevolution} (left).}.

One might think that if the sign problem is mild ($Z_+$ dominates), almost any method will succeed equally well. This is not necessarily the case, as we now demonstrate with the example of $\vert \det M \vert$. Note that an importance sampling method has to begin with a positive measure. In the case of multi-parameter reweighting we obtain this positivity by taking ${\rm Re}\,\mu_0 = 0$. In the $Z_{\rm new}$ method, we take $\vert {\rm Re} \det  M \vert$. A third possibility is $\vert \det M \vert$: in this method we sample from the measure $e^{-S_G (A)}  \vert \det M(A) \vert$ and treat the phase of $\det M(A)$ as part of the observable \cite{absdetMR1,absdetMR2}. However, this is not guaranteed to give a good approximation: it is possible that for typical configurations $\vert \det M(A) \vert$ is very different from $\vert {\rm Re} \det M(A) \vert$ (e.g., if the phase angle distribution is peaked near $\pm \pi / 2$). Sampling from $e^{-S_G (A)}  \vert \det M(A) \vert$ would then mostly obtain relatively atypical configurations (whereas $Z_{\rm new}$ would sample more typical ones), thus causing a kind of ``overlap problem'': of all reweighting-like methods, sampling with $Z_{\rm new}$ causes the smallest fluctuations in the associated reweighting factor \cite{rewfactor}, at least if one does not have any prior knowledge about the phase distribution of the fermion determinant or the actual severity of the sign problem. Note, there is no good notion of ``typicality'' in the first place if the sign problem is strong, since in this case all regions of the integration space are equally important in order to capture the fine cancellations between positive and negative contributions. The only thing a sampling method can possibly achieve in this case is to make fluctuations in the denominator and the numerator of the reweighting formula small \cite{rewfactor}. As a related point, in no reweighting method (lest ours) can the average of the reweighting factor (computed in the reweighted ensemble) be an absolutely reliable indicator for how well the reweighting method really captures the true ensemble (with the sign problem): namely, in the case where the reweighting factor is not small for the sole reason that the configurations on which one computes the average reweighting factor (and which have been sampled with the reweighted measure) are so untypical that they yield a relatively big (and therefore wrong) average reweighting factor although the same sampled configurations yield wrong observable averages otherwise; for this reason, one does not have a reliable indication as to when the reweighting method fails.

In terms of computational cost, our method probably does rely on full computation of ${\rm Re}\det M(A)$, as the sampling proceeds: since there is no apparent way of expressing ${\rm Re}\det M$ as a Gaussian (or, at least, exponential) bosonic integral (as can be done, e.g., for $\vert \det M\vert$ via pseudofermions \cite{Weingarten:1980hx}), algorithms that are based on Hamiltonian dynamics to update the entire lattice at once (such as Hybrid Monte Carlo (HMC) \cite{Duane:1987de}) do not seem applicable. Therefore, sampling from $\vert {\rm Re}\det M\vert$ instead seems to demand local link updates with evaluation of the non-local expression ${\rm Re}\det M$, requiring $\sim N^3$ operations with $N$ the dimension of the fermion matrix, for each of the $N$ Metropolis tests during one sweep through the lattice. By contrast, HMC sampling from $\vert \det M\vert$ needs less than $\sim N^{9/4}$ floating point operations to obtain a new decorrelated configuration. Nevertheless, even there $\det M$ has to be computed fully for each of the sampled, decorrelated configurations. Thus, in the end our method seems to be slower by a factor of at least $\sim N$ (depending on the autocorrelation properties of our method), which is a huge disadvantage for sensible lattice sizes. For small $\mu$, more efficient, but approximate, methods exist to determine the phase angle $\theta$ \cite{rewfactor,Ejiri:2004yw}, and thus ${\rm Re} \det M$. For example, one can expand $\det M$ in powers of $\mu$; each coefficient in the expansion is much less costly to evaluate than $\det M$ itself.

\bigskip

\emph{Acknowledgments ---} We thank Gert Aarts, Michael Endres, Simon Hands and especially Mark Alford, Phil\-ippe de Forcrand, Maria Paola Lombardo and Bob Sugar for useful comments. We are particularly grateful to Kim Splittorff for discussions concerning evidence for large sign fluctuations in different regions of the QCD phase diagram. The authors are supported by the Department of Energy under DE-FG02-96ER40969.

\end{document}